\documentstyle[bo99,epsfig]{article}

\title{What's wrong with AGN models for the X-ray background ?}
\author{Andrea Comastri}
\affil{Osservatorio Astronomico di Bologna, via Ranzani 1, 
I--40127 Bologna, Italy}                             

\begin{document}

\maketitle

\begin{abstract}

The origin of the hard X--ray background (XRB) as a superposition of
unabsorbed and absorbed Active Galactic Nuclei 
is now widely accepted as the standard model.
The identification of faint X--ray sources in 
{\sf ROSAT}, {\sf ASCA}, and {\sf BeppoSAX} medium--deep surveys 
and their average spectral properties are in broad agreement with 
the model predictions.
However, AGN models, at least in their simplified version, 
seem to be at odds with some of the most recent findings calling for
substantial revisions. 
I will review the recent XRB  
``best fit" models and discuss how the foreseen {\sf XMM} and {\sf Chandra}
surveys will be able to constrain the allowed parameter space.

               
\end{abstract}

\section{Introduction}

It has been recognized, already a few years ago, that a self--consistent
AGN model for the XRB requires the combined fit of several observational 
constraints in addition to the XRB spectral intensity 
such as the number counts, the redshift and absorption distribution  
in different energy ranges, the average spectra and so on
(Setti \& Woltjer 1989).     
First attempts towards a ``best fit" solution relied  
on simplified assumptions for the AGN  spectral properties
and for the evolution of their luminosity function
(Madau, Ghisellini \& Fabian 1994 (MGF94), Comastri et al. 1995 (CSZH95),
Celotti et al. 1995 (CFGM95)).
A three step approach has been followed to build the so--called
baseline model: the first step is to assume a single average spectrum for the
type 1 objects which is usually parameterized as a power law plus a reflection
component from a face--on disk and a high--energy cut--off at a few hundreds
of keV.
A distribution of absorbing column densities for type
2 objects is then added in the second step.
Finally the template spectra are folded with an evolving XLF
which, in the framework of unified models, 
does not depend on the amount of intrinsic obscuration.  
The number density and absorption distribution of obscured sources 
are then varied until a good fit is obtained.
The baseline model led to a successful description of most of the observational
data available before 1995 and to testable predictions for the
average properties of the sources responsible for the bulk of the XRB. 
The increasing amount of data from soft and hard X--ray surveys combined  
with the study of nearby bright sources has been used  
to obtain a more detailed description of the AGN X--ray spectra
and absorption distribution.
In addition, the optical identification of sizeable samples
of faint AGNs discovered in the {\sf ROSAT}, {\sf ASCA} and 
{\sf BeppoSAX} surveys
has shed new light on the evolution of the AGN luminosity
function opening the possibility to test in more detail the 
AGN synthesis model predictions. 
As a consequence, the modelling of the XRB has attracted renewed attention
and several variations/improvements with respect to the baseline model 
have been proposed. 
However, despite the increasing efforts, 
a coherent self--consistent picture of ``the" XRB model has yet to be reached,
as most of its ingredients have to be extrapolated well beyond 
the present limits.
Besides the interest in a best--fit model it is by now clear that the 
problem of the 
origin of the XRB is closely related to the  evolution of accretion
and obscuration in AGN.  
As a consequence, the XRB spectrum should be considered as a useful tool 
towards a better understanding of the history of black hole formation 
and evolution in the Universe (Fabian \& Iwasawa 1999) and the 
interplay between AGN activity and star--formation (Franceschini et al. 1999;
Fabian this volume).

\section{Recent Observational constraints}

\subsection{The XRB spectrum}

The low energy (below 10 keV) XRB spectrum has been measured  
with the imaging detectors onboard {\sf ROSAT}, {\sf ASCA}, and {\sf BeppoSAX}
and a summary of the results is given in Figure 1 together 
with a compilation of recent re--analysis of the {\sf HEAO1 A2} and 
{\sf A4} experiments
data. The comparison between the different datasets in the overlapping 
$\sim$ 1--8 keV energy range points to a systematic difference
in the normalization of the XRB flux while the average spectrum 
is similar among all the observations. 
The largest deviation is of the order of $\sim$ 40 \%
between the {\sf HEAO1 A2} and {\sf BeppoSAX data} 
(see Vecchi et al. 1999 for a more detailed discussion).
Such a discrepancy could be due to residual, not fully understood, 
cross--calibration errors among the different detectors 
and/or to field--to--field fluctuations. 
These findings cast shadows on the intensity and the location of the XRB peak
as measured by {\sf HEAO1 A2} ($\sim$ 43 keV cm$^{-2}$ s$^{-1}$ sr$^{-1}$ at
30 keV; Gruber et al. 1999).
Indeed a simple extrapolation of the {\sf BeppoSAX} and {\sf HEAO1 A4 MED} 
best fit spectra imply that the XRB spectrum peaks at $\sim$ 23 keV with
a much higher intensity introducing an extra--degree of freedom  
in AGN models parameter space.
A new measurement of the 10--100 keV spectrum would be extremely important.
Unfortunately such observations are not foreseen in the near future.

\subsection{The AGN spectrum}

As far as the model of the XRB is concerned, the most important parameters 
to deal with are a good estimate of the average continuum slope 
and of the absorption column density.
The broad--band energy range exploited by {\sf BeppoSAX} turned out to be 
extremely useful to probe column densities as high as 
10$^{24-25}$ cm$^{-2}$, to assess
the strength of the reflection component which peaks around 20--30 keV, and
the shape of the low--energy soft--excess emission below $\sim$ 1 keV. 
In addition {\sf ASCA} observations of sizeable samples of relatively
faint AGNs have allowed to probe the spectral properties of
high--luminosity high--redshift objects. 
The most important new results emerging from these observations
can be summarized as follows:

\begin{figure}
\centerline{\psfig{file=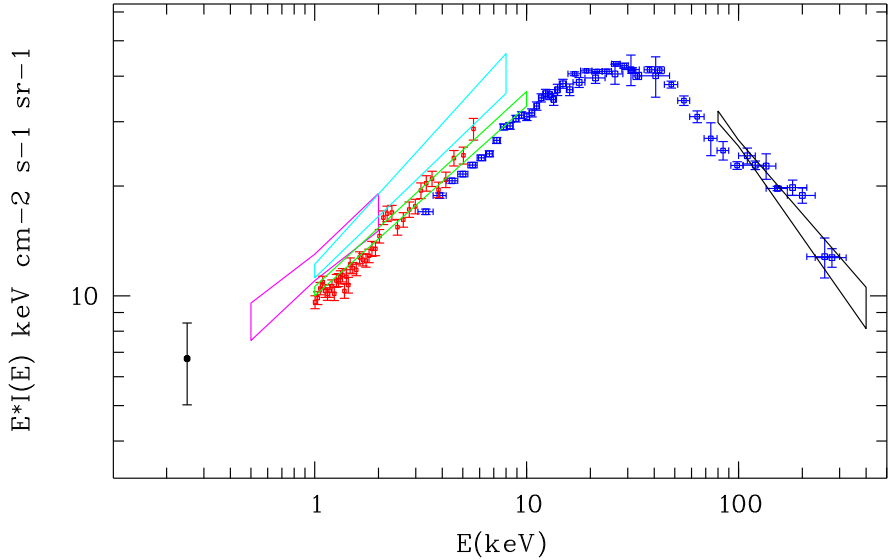, width=9cm, angle=-90}}
{\vskip -1cm\caption[]{The XRB spectral energy density from 0.25 to 400 keV.
The red points in the 1-6 keV range are
from {\sf ASCA} (Gendreau et al. 1995), the blue points
in the 3--300 keV range are from {\sf HEAO1 A2} and {\sf A4 LED} experiments
(Gruber et al. 1999), while the best-fit spectrum
from the {\sf HEAO1 A4 MED} experiment (Kinzer et al. 1997) is reported
as a black bow--tie contour between 80 and 400 keV.
The best-fit estimates from recent observations at low energies ($<$ 10 keV)
are also displayed as bow-ties contours. Green (1-10 keV):
 joint {\sf ROSAT/ASCA} analysis (Miyaji et al. 1998). Purple (0.5-2.0 keV):
{\sf ROSAT} results from Georgantopoulos et al. (1996). Cyan (1-8 keV):
 {\sf BeppoSAX} data from Vecchi et al. (1999).
Finally an estimate of the
extragalactic background intensity at 0.25 keV (Roberts \& Warwick 1998)
is also plotted.}
}
\end{figure}

\par\noindent
$\bullet$ The fraction of heavily obscured (24 $<$ log $N_H$ $<$ 25) and 
Compton thick (log $N_H >$ 25) sources in the local Universe is
much higher than previously thought (Risaliti, Maiolino \& Salvati 1999)
and a fraction as high as 50\% of the Seyfert 2 in the 
local Universe could be obscured by these large column densities.

\par\noindent
$\bullet$ Soft excess emission is uncommon among bright
Seyfert 1 galaxies (Matt this volume) and nearby quasars
(George et al. 2000) and estimated to be present in less 
than $\sim$ 30 \% of AGN.

\par\noindent
$\bullet$ First observations of high redshift quasars suggest
a flattening of the power law slope which cannot be ascribed 
to the reflection component (Vignali et al. 1999). 

\par\noindent
$\bullet$ Despite intensive searches for high luminosity highly absorbed
objects (the so--called type 2 quasars) these sources appear to be elusive
and only a few bona--fide examples have been reported in the 
literature (i.e. Barcons et al. 1998; Georgantopoulos et al. 1999).

\subsection{The evolution of the AGN X-ray luminosity function}

The evolution of the AGN XLF 
has been extensively studied mainly in the soft X--rays 
and usually parametrized with a pure luminosity evolution
({\rm PLE}) model (i.e. Boyle et al 1994). 
A major step forward in the determination of the soft XLF 
has been recently achieved by Miyaji et al. (2000).
Combining the results of several {\sf ROSAT} surveys
it has been possible to explore the low-luminosity high-redshift
tail of the XLF in much greater detail than before.
The results favour a luminosity dependent density evolution
({\rm LDDE}) as the best description of the available data.
In agreement with previous studies, X-ray selected AGN undergo
strong evolution up to a redshift $z_c$ = 1.5--2.0 and a levelling--off 
or a weak negative evolution up to $z_{max} \simeq$ 4--5.
Two parametric descriptions 
({\rm LDDE1} and {\rm LDDE2}) encompassing the 
statistically acceptable fits to the soft XLF have been worked out
by Miyaji and collaborators. 
The integration of the {\rm LDDE1} and {\rm LDDE2} XLF up to $z \simeq$ 5
accounts for about 60 \% and 90 \%  of the soft XRB respectively.
  
\section{The AGN models parameter space}

\subsection {Warnings}

Before discussing and comparing the various models, 
it is important to stress the strong coupling between 
the input spectral parameters and those describing the XLF evolution,
which instead are often uncorrectly considered to be independent 
in the models.
Indeed the X--ray luminosities are usually computed 
converting count rates into fluxes assuming a single
valued (relatively steep) slope.
This procedure might easily lead to
a wrong estimate of the intrinsic luminosity for a very hard absorbed 
spectrum or if the soft X--ray flux is due to a
component not directly related to the obscured nucleus
(as in the case of a thermal spectrum from a starburst or scattered 
emission).
According with the XRB baseline model, absorbed AGN 
become progressively more important towards faint fluxes and
thus an additional spurious density evolution term can be introduced 
in the derivation of the XLF. 
It turns out that not only the evolution and the space density of 
obscured AGN are highly uncertain, but also 
the common practice to consider the soft XLF  
as representative of the properties of type 1 objects 
is likely to contain major uncertainties especially when 
extrapolated to higher energies.
Unfortunately our present knowledge of
the AGN spectral and evolutive properties does not allow
to disentangle the spectral and evolutionary parameters, leaving
this ambiguity in all the XRB synthesis models.

\subsection{An incomplete tour of the parameter space}

The baseline model (cfr $\S$ 1) has been recently extended, 
taking into account some of the new observational findings 
described in $\S$2, by several authors:  
Gilli, Risaliti \& Salvati 1999 (GRS99); Miyaji, Hasinger \& Schmidt
 1999 (MHS99); Wilman \& Fabian 1999 (WF99); 
Pompilio, La Franca \& Matt 2000 (PLM00). 
A good agreement among the various models 
has been reached on the high energy cut--off in the input spectrum
(300--500 keV), which is basically fixed by the XRB shape above 40 keV
(Comastri 1999), and on the $z_c$ and $z_{max}$ values.
GRS99 and MHS99 adopted the {\rm LDDE} model for the evolution of 
the XLF and also introduced a cut--off in the luminosity 
distribution of absorbed AGN for $L >$ 10$^{44}$ erg s$^{-1}$ to cope
with the lack of type 2 QSO. The absorption distribution 
has been fixed according to the recent {\sf BeppoSAX}
results only in the GRS99 model.
PLM00 and WF99 both stressed that a proper treatement 
of the high energy spectrum of heavily obscured (24 $<$ log$N_H$ $<$ 25) 
objects has important consequences for the modelling.
In particular the evolution of the obscured to unobscured ratio 
as a function of redshift (PLM00) or the need of super--solar abundances
to better fit the XRB peak at 30 keV (WF99; but see $\S$ 2.1) have been invoked.

\begin{table}[h]
\centerline{\bf Table.~1 - Comparison of model parameters}
\begin{center}
\begin{tabular}{|l|c|c|c|c|c|c|c|}
\hline
\multicolumn{1}{l}{\bf Model} &
\multicolumn{1}{c}{\bf XLF$^a$} &
\multicolumn{1}{c}{\bf Evolution} &
\multicolumn{1}{c}{\bf QSO2$^b$ } &  
\multicolumn{1}{c}{\bf $N_{\rm H}^c$} &
\multicolumn{1}{c}{\bf $\alpha_E^d$} &
\multicolumn{1}{c}{\bf SE$^e$} &
\multicolumn{1}{c}{\bf CT$^f$} \\
\hline
{\bf MGF 94} & 2-10  & PLE & Yes & Fitted & 0.9 & No & Yes   \\
{\bf CSZH 95} & 0.3-3.5 & PLE & Yes & Fitted & 0.9 & Yes & Yes(*) \\
{\bf CFGM 95} & 2-10  & PLE & Yes & Fitted & 0.9 & No & Yes \\
{\bf MHS 99} & 0.5-2.0 & LDDE1 & No & Fitted & 0.7 & Yes & No  \\
{\bf GRS 99} & 0.5-2.0 & LDDE1 & No & Fixed & 0.9 & Yes & Yes(*)   \\
{\bf WF 99}   & 2-10 & PLE & Yes & Fitted & 0.9 & No & Yes  \\
{\bf PLM 00}   & 0.3-3.5 & PLE & Yes & Fitted & 0.9 & Yes/No & Yes   \\
\hline
\end{tabular}
\end{center}
\vspace{-0.2cm}
\hspace{0.1cm} {$^{a}$ Energy range of the adopted XLF;
 $^{b}$ Presence of type 2 quasars; $^{c}$ Absorption distribution; 
$^{d}$ Spectral energy slope; 
$^{e}$ Presence of soft excess emission in the 
model spectrum; $^{f}$ Presence of Compton thick sources 
(The * indicates that a simplified treatment 
has been employed)
}
\end{table}

A comparison between the various models (all of them providing
a fairly good description of the present data) is made difficult
by the large dispersion in the starting assumptions among the
different authors (see Table 1) and also by the relatively large
uncertainty in the XRB spectrum normalization (see $\S$ 2.1).

\noindent
The most up-to-date treatment of the XLF evolution has been adopted only by 
GRS99 and MHS99 who also made an attempt to correct for the biases 
described in $\S$ 3.1. 
In both cases the model predictions fall short the
hard X--ray (2--10 keV and especially 5--10 keV) counts at relatively
bright 10$^{-13}$--10$^{-12}$ cgs fluxes.  
This effect, which is less severe for MHS99 given the very hard
input spectrum ($\alpha$ = 0.7 plus reflection), 
can be explained by the relatively low average luminosity of absorbed
sources which show up only at fainter fluxes.
The hard X--ray counts are better accounted for in PLE  
models (Fig.~2), which however are 
based on a less appropriate description of the XLF and include 
high luminosity highly absorbed sources.
It is worth noting that source counts at fluxes $>$ 10$^{-13}$ cgs,
both in the soft and hard bands, 
should not be entirely accounted for by AGN models as a non negligble
fraction of these relatively bright sources are not AGN.
The 2--10 keV and 5--10 keV counts are best fitted by those
models without soft excess emission in type 1 objects.
However in this case the predicted average spectrum of faint sources
in the {\sf ROSAT} band ($\alpha_E \simeq$ 0.5--0.6) is much harder than the
observed value ($\alpha_E \simeq$ 1.0, Hasinger et al. 1993).

\begin{figure}
\centerline{\psfig{file=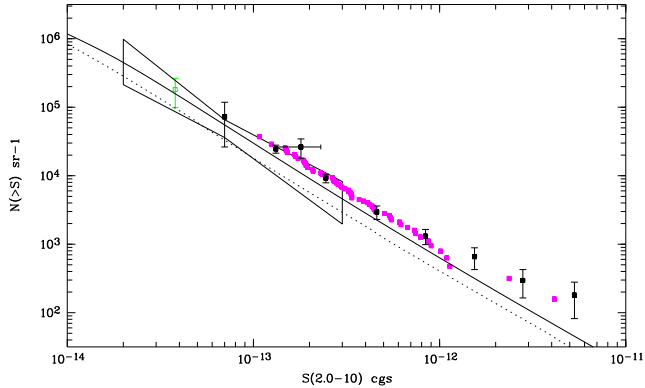, width=6.5cm, angle=-90}}
{\vskip -1cm\caption[]{The CSZH95 (solid line) and the GRS99 
(dotted line) AGN model predictions are compared with the 
total 2--10 keV counts. 
Data points at bright fluxes are from Cagnoni et al. 1998 
(single sources plotted); Ueda et al. 1999 (points with error bars); 
Ogasaka et al. 1998 (at the faintest flux). 
The bow--tie contour is from a fluctuation analysis 
of {\sf ASCA} data by Gendreau, Barcons \& Fabian 1998.} 
}
\end{figure}

Another inconsistency of most of these models concerns the relatively 
small expected percentage of  
type 1 unobscured AGN at the 2--10 keV fluxes currently sampled. 
Indeed optical identifications of medium--deep 
{\sf ASCA} surveys (Boyle et al. 1998; Akiyama et al. 2000) 
suggest that the fraction of unabsorbed broad line AGN is of the 
order of 60--70 \% while only one third of the sources should be type 1 AGN 
on the basis of the models predictions (but see $\S$ 4).
The fraction of type 1 AGN  can be increased assuming
an LDDE2 model for the evolution of the XLF and 
a flat $\alpha_E$ = 0.7  spectrum for high luminosity objects
(Vignali et al. 1999). With these parameters a 
good fit to the hard XRB spectrum can be obtained 
even without including heavily obscured ($N_H >$ 10$^{24}$) sources.
As a result the relative ratio between absorbed 
and unabsorbed objects at relatively bright fluxes (Fig.~3)
decreases significantly. 
However also within this model the hard X--ray counts are 
seriously underestimated (being consistent with the dotted line in Fig.~2)
owing to the decreased emissivity of hard absorbed sources.

\begin{figure}
\centerline{\psfig{file=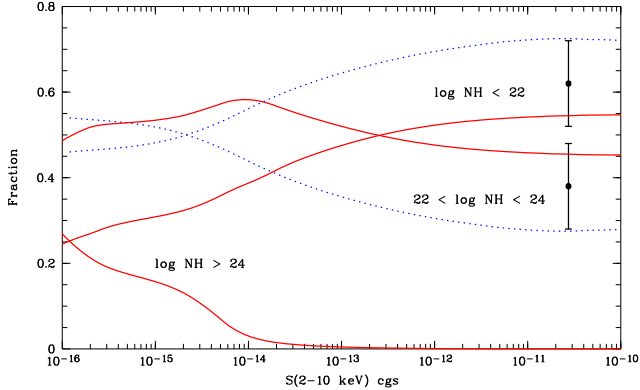, width=6.5cm, angle=-90}}
{\vskip -1cm\caption[]{The relative fraction of unobscured 
and obscured sources for a LDDE2 model (dotted line) and
a PLE model (solid line). The points with error bars represent the fraction
of unabsorbed (about 60\%) and absorbed (about 40\%)  
sources in the Piccinotti et al. (1982) sample.}
}
\end{figure}

As an example of the link between the various parameters 
I have computed three different models which 
provide a good fit to the overall XRB spectrum but differ in 
the choice of the input spectra and XLF evolution (Fig.~4).
Assuming a high fraction of type 1 objects 
as in the LDDE2 scenario a soft excess component cannot be accomodated 
as the 1/4 keV background would be overpredicted.
On the other hand the class of PLE models without soft excess
(which better reproduce the hard X--ray counts, but 
with the caveats discussed above)
suggest a possible contribution from 
other, steep spectrum, sources to the 0.25 keV background.

\begin{figure}
\centerline{\psfig{file=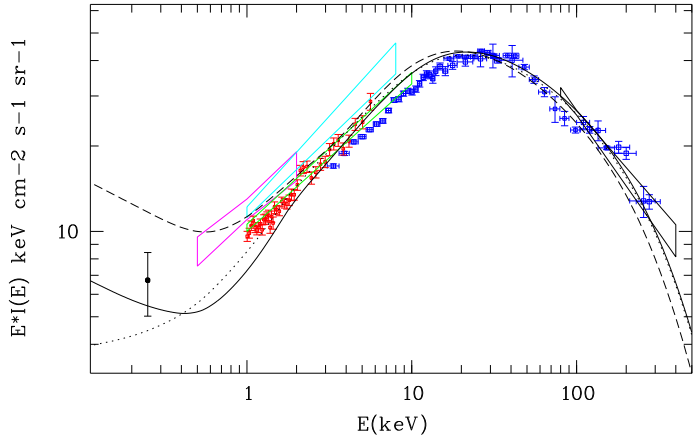, width=7cm, angle=-90}}
{\vskip -1cm\caption[]{XRB fits normalized at 30 keV for 3 different assumptions on the
input spectral shape and XLF evolution. Solid line : PLE with Soft Excess. 
Dotted line: PLE without soft excess. Dashed line: LDDE2 with soft excess.
The observational data are the same of Fig. 1.}
}
\end{figure}

\section{Is there a way out ?}  

The main message emerging from what discussed above is that a 
self--consistent description of all the observational constraints 
is still lacking. 
The major problem is the discrepancy between the predictions of 
those models computed assuming the most up--to--date results, 
and the high energy ($>$ 2 keV) source counts.
One obvious possibility is a substantial contribution 
from non--AGN, flat spectrum sources.  
Extremely hard ($\alpha_E \simeq$ 0.2) power--law tails 
above a few keV, possibly originating in advection dominated accretion flows,
have been recently discovered in a small sample of nearby elliptical 
galaxies (Allen, Di Matteo \& Fabian 2000).
It has been proposed (Di Matteo \& Allen 1999) that these objects 
constitute the missing population needed to 
fill the gap between the hard counts and the AGN model predictions. 
However in this case elliptical galaxies should be a non--negligible   
fraction of the already identified X--ray sources in
{\sf ASCA} and {\sf BeppoSAX} surveys at variance with the 
present breakdown of optical identifications. 

Another interesting possibility, which would allow to 
include high luminosity highly absorbed AGN in the models 
and at the same time reproduce most of the observational constraints,
is that the optical properties of X--ray obscured AGN are different 
from what expected (i.e. narrow lined AGN).
In this respect the identification of the first High Energy LLarge
Area Survey (HELLAS) carried out with {\sf BeppoSAX} in the 5--10 
keV band (Fiore et al. 1999, and this volume) is providing 
new and unexpected results. In particular, X--ray absorbed AGN are 
identified with objects which show a large variety of optical classification, 
such as 
intermediate type 1.5--1.9 objects, red quasars (Vignali et al. 2000)
and even broad line ``blue" quasars. A similar behaviour has been
also reported for a sample of {\sf ROSAT} AGN 
(Mittaz, Carrera, Page this volume). 
It is also interesting to note that large columns of cold 
gas have been detected in Broad Absorption Line quasars
(Brandt et al. this volume) and in several Broad Line Radio galaxies
and radio quasars observed by {\sf ASCA} 
(Sambruna, Eracleous \& Mushotzky 1999).
Although the statistics is not yet good enough to reach firm 
conclusions, it is quite possible that the correlation between 
X--ray absorption and optical appearance of AGN  
change with redshift and/or luminosity (Fig.~5).
A decreasing value of the dust--to--gas ratio as a function of 
the X--ray luminosity would provide a possible explanation of this effect.


\begin{figure}[ht]
\centerline{
\hbox{
\psfig{file=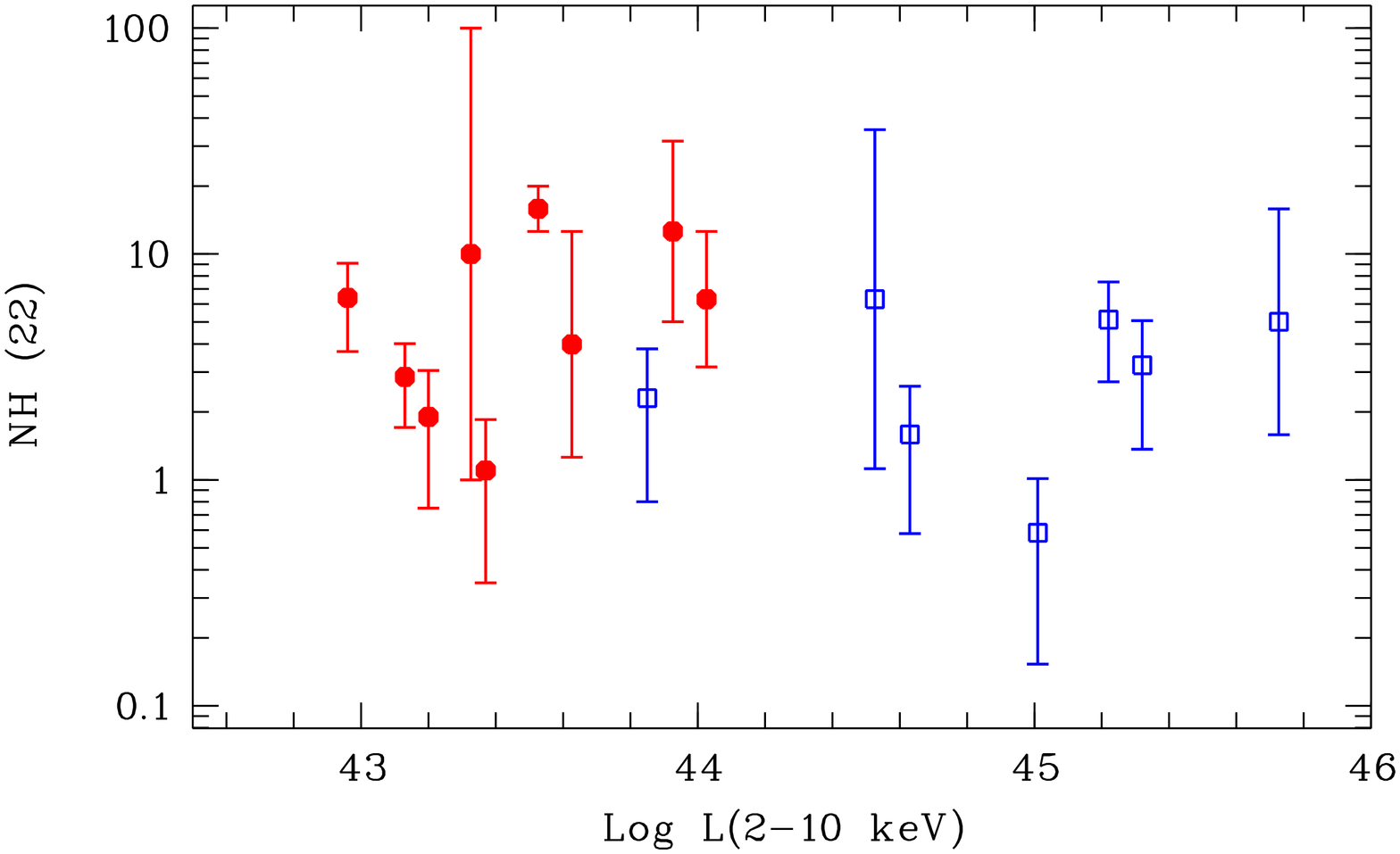, width=5cm, angle=-90}
\psfig{file=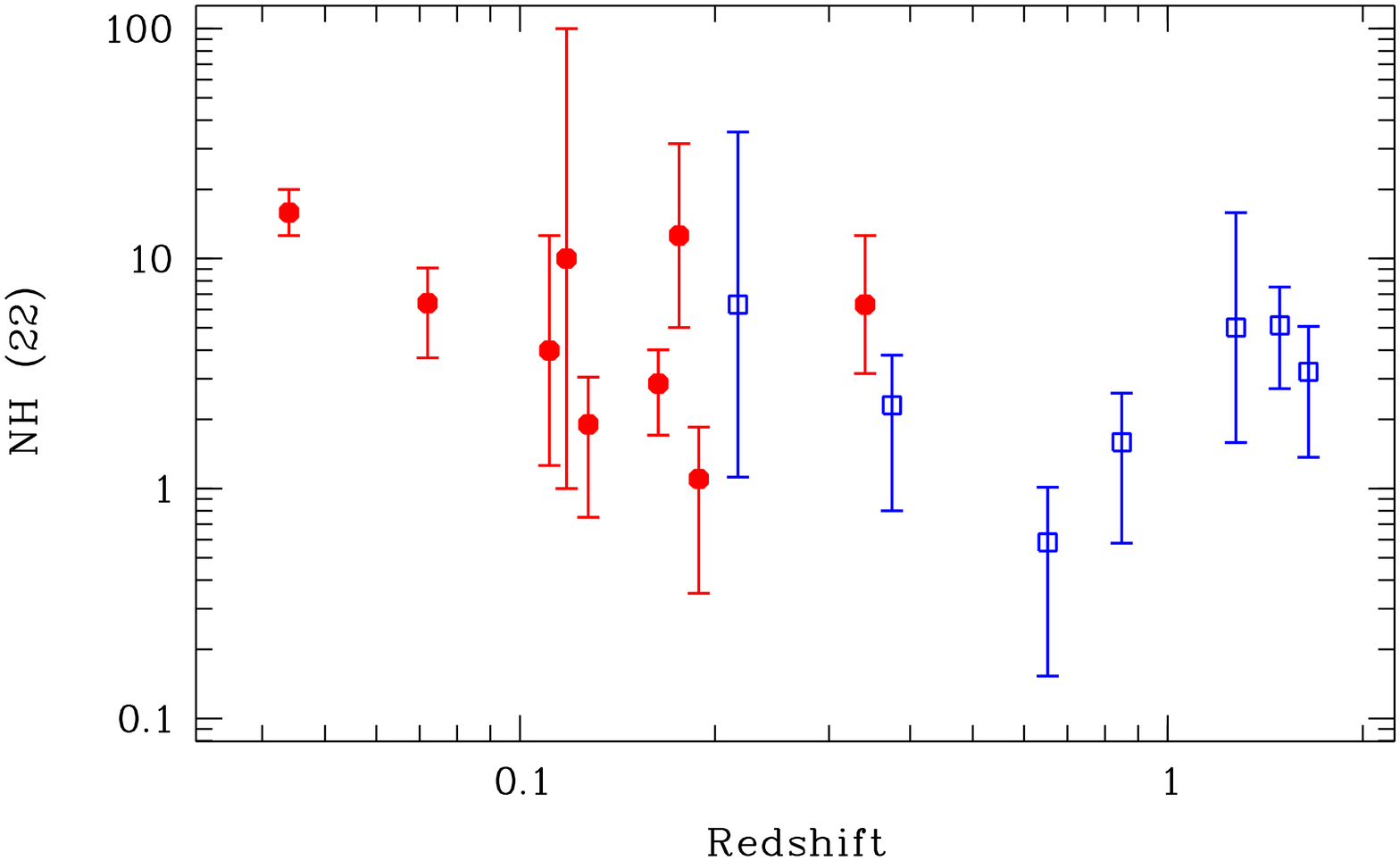, width=5cm, angle=-90}
}
}
\caption{The absorption column density (in units of 10$^{22}$ cm$^{-2}$
versus the 2-10 keV luminosity (left panel) and redshift (right panel) 
of obscured AGN in hard X-ray selected samples. Broad lined objects are 
plotted with open (blue) symbols, narrow lined ones with filled (red) symbols.}
\label{zlx}
\end{figure}

\section{Conclusions}

In order to achieve a major improvement in the exploration of 
XRB models parameter space, the resolved fraction 
of its energy density should be of the order of 
50--60 \% or higher. 
\begin{table}[h]
\centerline{\bf Table.~2 - Resolved fraction of the XRB}
\begin{center}
\begin{tabular}{|c|c|c|}
\hline
\multicolumn{1}{l}{\bf 2-10 keV Flux interval} &
\multicolumn{1}{c}{\bf Relative \%} &
\multicolumn{1}{c}{\bf Integral \%} \\
\hline
 $>$ 10$^{-11}$       & 0.5  & 0.5  \\
 10$^{-12}$-10$^{-11}$ & 2 & 2.5 \\
 10$^{-13}$-10$^{-12}$ & 8 & 10.5  \\
 10$^{-14}$-10$^{-13}$ & 32.5 & 43  \\
 10$^{-15}$-10$^{-14}$ & 39 & 82  \\
 10$^{-16}$-10$^{-15}$ & 16 & 98  \\
 $<$ 10$^{-16}$       & 2 &  100 \\
\hline
\end{tabular}
\end{center}
\end{table}
The expected contribution of AGN 
to the 2--10 keV XRB is reported in Table 2 as a function of flux. 
The model parameters
are such to account for an intensity of
$\sim$ 7 $\times$ 10$^{-8}$ erg cm$^{-2}$ s$^{-1}$ sr$^{-1}$
(in between the {\sf ASCA} and {\sf BeppoSAX} measurements) 
at $\sim$ 10$^{-17}$ cgs.  
The predictions are model dependent and should be considered as indicative.
Nevertheless it is clear that at the fluxes sampled by the  
foreseen {\sf Chandra} and {\sf XMM} medium--deep surveys 
most of the XRB will be resolved allowing to 
unveil the nature of the sources making the bulk of its energy density.
The most important challenge for XRB models will be 
the study of X--ray absorption and luminosity distribution 
for 2--10 keV fluxes $<$ 10$^{-13}$ cgs, the search for 
heavily obscured AGN which according to the predictions are expected 
to show up in a substantial fraction below $<$ 10$^{-14}$ cgs (cfr. Fig.~3),
and the optical--infrared follow--up of X--ray obscured sources.

\begin{acknowledgements}
Partial support from ASI contract ARS-98-119 and MURST grant 
Cofin98-02-32 is acknowledeged. I thank G. Zamorani and R. Gilli 
for useful discussions.
\end{acknowledgements}


\begin{references}

\ref{} Allen S.W., Di Matteo T., Fabian A.C., 2000, MNRAS 311, 493
  
\ref{} Akiyama M., et al., 2000, ApJ in press (astro-ph/0001289)
 
\ref{} Barcons X., et al., 1998, MNRAS 301, L25
 
\ref{} Boyle B.J., et al., 1994, MNRAS 271, 639
  
\ref{} Boyle B.J., et al., 1998, MNRAS 296, 1

\ref{} Cagnoni I., Della Ceca R., Maccacaro T., 1998, ApJ 493, 54

\ref{} Celotti A., Fabian A.C., Ghisellini G., Madau P., 1995, MNRAS 277, 1169   

\ref{} Comastri A., Setti G., Zamorani G., Hasinger G., 1995, A\&A 296, 1

\ref{} Comastri A., 1999, Astr. Lett. \& Comm. 39, 181

\ref{} Di Matteo T., Allen S.W., 1999, ApJ 527, L21 

\ref{} Fabian A.C., Iwasawa K., 1999, MNRAS 303, L34  

\ref{} Fiore F., et al., 1999, MNRAS 306, L55

\ref{} Franceschini A., et al., 1999, MNRAS 310, L5

\ref{} Gendreau K.C., et al., 1995, PASJ 47, 5

\ref{} Gendreau K.C., Barcons X., Fabian A.C., 1998, MNRAS 297, 41

\ref{} Georgantopoulos I., et al., 1996, MNRAS 280, 276 

\ref{} Georgantopoulos I., et al., 1999, MNRAS 305, 125  

\ref{} George I.M., et al., 2000, ApJ 531, 52 

\ref{} Gilli R., Risaliti G., Salvati M., 1999, A\&A 347, 424

\ref{} Gruber D.E., et al., 1999, ApJ 520, 124  

\ref{} Hasinger G., et al., 1993, A\&A 275, 1

\ref{} Kinzer R.L., et al.,  1997, ApJ 475, 361 

\ref{} Madau P., Ghisellini G., Fabian A.C., 1994, MNRAS 270, L17

\ref{} Miyaji T., et al. 1998, A\&A 334, L13 

\ref{} Miyaji T., Hasinger G., Schmidt M., 1999, Adv. Space Res. in press

\ref{} Miyaji T., Hasinger G., Schmidt M., 2000, A\&A 353, 25

\ref{} Ogasaka Y., et al., 1998, AN 319, 43

\ref{} Piccinotti G., et al., 1982, ApJ 253, 485 

\ref{} Pompilio F., La Franca F., Matt G., 2000, A\&A 353, 440 

\ref{} Risaliti G., Maiolino R., Salvati M., 1999, ApJ 522, 157 

\ref{} Roberts T.P., Warwick R.S., 1998, AN 319, 34

\ref{} Sambruna R.M., Eracleous M., Mushotzky R.F., 1999, ApJ 526, 60 

\ref{} Setti G., Woltjer L., 1989, A\&A 224, L21

\ref{} Ueda Y., et al., 1999, ApJ 524, L11 

\ref{} Vecchi A., Molendi S., Guainazzi M., et al., 1999, A\&A 349, L73  

\ref{} Vignali C., Comastri A., Cappi M., et al., 1999, ApJ 516, 590 

\ref{} Vignali C., Mignoli M., Comastri A., et al., 2000, MNRAS in press (astro-ph/0002279) 

\ref{} Wilman R.J., Fabian A.C., 1999, MNRAS 309, 862

\end{references}
\end{document}